# A MUON SOURCE PROTON DRIVER AT JPARC-BASED PARAMETERS*

D. Neuffer[#], Fermilab, Batavia, IL 60510, USA

*Abstract*

An "ultimate" high intensity proton source for neutrino factories and/or muon colliders was projected to be a ~4 MW multi-GeV proton source providing short, intense proton pulses at ~15 Hz. The JPARC ~1 MW accelerators provide beam at parameters that in many respects overlap these goals. Proton pulses from the JPARC Main Ring can readily meet the pulsed intensity goals. We explore these parameters, describing the overlap and consider extensions that may take a JPARC-like facility toward this "ultimate" source. JPARC itself could serve as a stage 1 source for such a facility.

## INTRODUCTION

Each Future applications using high intensity muon beams, such as a neutrino factory or a muon collider will require a very high power proton source. The IDS neutrino factory study specified a 4 MW proton source operating at ~60 Hz.[1] Muon Collider scenarios require ~4 MW sources providing intense pulses of protons at ~15 Hz. [2]

All of these designs require a new proton source in the multi-MW class. At Fermilab, a proton source based on a 4 MW 8 GeV proton linac was considered. The 8 GeV linac matched Fermilab's need for a linac/booster replacement to obtain multi-MW beams for long baseline neutrino facilities. The linac itself could provide the beam power but not the time structure of the ultimate source. The Fermilab MAP plan was to accumulate beam at 8 GeV in an accumulator/compressor ring.[3] and the accumulated beam could be bunched in accumulator/compressor ring(s) to obtain the desired time structure. Parameters for the accumulator/compressor are included in Table 1. The accumulator would provide 8 GeV beam at 15 Hz in 4 bunches. The bunches could be compressed and extracted one at a time to provide 60 Hz pulses or 4 bunches could be extracted and simultaneously compressed onto a target for 15Hz pulses. Other muon source designs are presented in ref. 4.

These scenarios are constrained by space charge forces, and the strength of the space charge force in a ring can be estimated by using the formula for tune shift $\Delta \nu$:

$$\Delta \nu = \frac{3 N_p r_p}{2 \varepsilon_{N,95} \beta \gamma^2 B_F} \quad [1]$$

where $N_p$ is the number of protons in the ring, $r_p$ =1.838×10$^{-18}$ m, $\varepsilon_{N,95}$ is the 95% normalized emittance ($\varepsilon_{N,95} = 6\pi \varepsilon_{N,rms}$), and $B_F$ is the bunching factor. (This overly simplified formula ignores the dispersion-dependent beam size ($\eta \delta p/p$), x-y asymmetry, and assumes gaussian shape.) At typical Fermilab Booster injection parameters ($N_p$= 4.2×10$^{12}$, E=400 MeV, $B_F$ =1/3, $\varepsilon_{N,95}$=15π mm-mrad), $\Delta \nu$ =~0.42. In general, $\Delta \nu$ should be significantly less than 1, and 0.4 is somewhat undesirably large. (The oversimplification has increased the value.)

The 8 GeV linac was beyond the near-term cost constraints imposed by DoE. In the short term, a 0.8 GeV linac (PIP-II) upgrade that meets DoE cost goals has received CD-0 approval. This could be upgraded to an 8 GeV linac to meet performance goals, but this may not be the optimum choice within DoE cost constraints.

## MUON SOURCE AND JPARC ACCELERATORS AND PARAMETERS

The parameters of the JPARC accelerators have evolved somewhat in construction and initial operation, and have some potential upgrade paths. The core components are a 400 MeV linac, a 3 GeV rapid-cycling synchrotron (RCS), and a 30---50 GeV main ring (MR).[5] The operation of the facility has been delayed and restricted by the Japan earthquake and other difficulties, but is now approaching its design goals. The RCS and MR are designed to reach ~1 MW power levels, and the RCS has already done that[6, 7]. Some parameters of these accelerators are presented in Table 1, and compared to the MAP example. Both rings use relatively low frequency rf, and thus can readily provide the single high-intensity short bunches needed for the production and capture scenarios of the collider.

The RCS has recently operated at 25 Hz with 8.4×10$^{13}$ particles per pulse (4.2×10$^{13}$ per bunch), obtaining 1 MW operation. The bunch and pulse intensity parameters, and time structure, meet the goals of the ultrahigh intensity muon source (UMS). The intensities are more than an order of magnitude larger than those achieved in the Fermilab Booster. The highest intensity operation requires careful painting in transverse and longitudinal phase space at injection, nominally to 100π mm-mrad emittance (This is a 90% emittance ($4\pi\sigma^2/\beta_t$), corresponding to 150π at Tevatron 95% units ($6\pi\sigma^2/\beta_t$).).

Use of eq. 1 with $B_F$ = 0.4, obtains $\Delta \nu$ = 0.71. The tune shift is actually ~0.4. (Painting has flattened the distribution by a factor of ~2.) The intensity per bunch is similar to that of the UMS source (4.2 rather than 5.2 ×10$^{13}$), and the time structure (25 Hz with bunches at 50 Hz) is similar. This similarity strongly indicates that a proton source at UMS parameters is credible and operable and would require only modest extension of the RCS design.

In the initial design parameters for the MR, JPARC planned to produce 50 GeV beam at 0.3 Hz and 3.3×10$^{14}$ p/pulse to deliver ~0.8 MW beam power. The MR can reach 50 GeV, but 30 GeV operation requires much less magnet power and is used in current operation. It has operated at up to 0.35 MW and in intensity upgrade

*Work supported by FRA Associates, LLC under DOE contract DE-AC02-07CH11359
[#]neuffer@fnal.gov

studies at up to $3.4 \times 10^{13}$ p/bunch. With $3.4 \times 10^{13}$ p/bunch, 8 bunches at 30 GeV, and 0.4 Hz operation, beam power would reach 0.52 MW and that would be a near-term operational goal. From that point 1 MW could be reached by reducing the cycle time to 1.25s. (0.8 Hz) An rf upgrade to enable this rep rate will be installed. Intensity per bunch could be increased to that of the RCS ($4.2 \times 10^{13}$), which would provide 1.3 MW with 0.8 Hz operation. (1.6 MW at 1Hz…)

At these higher pulse intensities, space charge is also an important consideration for the MR. At $3.3 \times 10^{14}$ p/pulse, $B_F = 0.4$, $E_p$ =3GeV injection, and $\varepsilon_{N,95}$= 100π μ, eq. 1 gives $\Delta \nu = 0.35$.

With $4.2 \times 10^{13}$ /bunch, a 30 GeV beam bunch contains 0.2MJ (0.33MJ at 50 GeV). A 60 Hz 4MW UMS source has only 0.067 MJ/pulse (a 15 Hz source has 0.27MJ/pulse). Thus, individual JPARC bunches will be as intense in total energy as the desired intensity of the UMS source. The phase space of a bunch is ~3 eV-s (full phase-space area) in the present painting injection into the RCS, and that phase space could be compressed within a 1m rms bunch length (full width 4 m), with appropriate rf. Thus the JPARC Main Ring readily produces beam pulses of the intensity needed for MAP-like muon production scenarios, and the JPARC Main Ring could be readily modified to provide the compressed bunches on target for a neutrino factory. A collider could also be initiated, but luminosity would be a bit less than presently desired, unless power is increased by a factor of ~2—4. (The 30—50 GeV protons are expected to be less efficient at π production than 8 GeV protons, by a factor of ~30%/MW, which may increase power requirements.)

Key components of the RCS and MR designs that enable high intensity include large apertures (12.5 cm radius in RCS and ~7cm in MR). The emittance acceptance is at least an order of magnitude greater than the Fermilab Booster. (The Fermilab Booster has an acceptance of ~±2cm.) Both RCS and MR lattices also avoid crossing transition in operation.

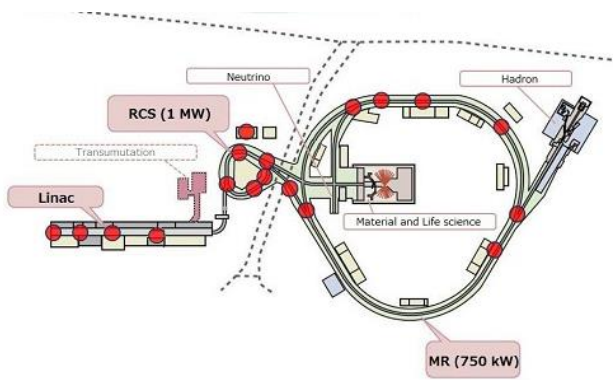

Figure 1: Overview of JPARC facility, showing locations of the Linac, 3 GeV RCS and 50 GeV Main ring.

## COMMENTS ON SCENARIO

One somewhat similar system is the Fermilab PIP→PIP2→PIP-N upgrade program, which will include a high-intensity MW-scale linac, and initially use the existing 8 GeV Booster to obtain 8 GeV beam at ~0.1 MW and then use the Booster to feed the Main Injector to obtain ~1MW at 60---120 GeV. Space charge within the existing Booster is believed to limit that upgrade to the ~1MW level. Higher-intensity would require a higher energy linac injector and/or a replacement for the Booster. The JPARC example indicates that an RCS replacement for the Booster could reach much higher intensities, but, to reach these much higher intensities, the RCS would require a much larger aperture as well as avoiding transition.

However, the Booster replacement (BR) is constrained by the requirement that its output beam will be used for injection into the existing Recycler and Main Injector, and the emittance acceptance of these rings is relatively small. The output beam must fit into the acceptance of the Main Injector (~40π mm-mrad normalized), which is almost an order of magnitude smaller than those rings. A 40π acceptance 0.8→8 GeV BR producing a 2 MW beam in the Main Injector requires ~2.5*$10^{13}$ p/pulse and would have a space charge tune shift of ~0.4 at BR injection. If operable at 20 Hz, the 8 GeV ring would be a 0.25MW source, substantially less than the JPARC machines. These values could be increased if the injection energy of the BR were increased (~2 GeV has been considered) and/or if the extraction energy were increased (~20 GeV). Intensity increase for the BR itself could be accommodated if it had larger apertures (>100π) but could be operated with the smaller beam sizes (<40π) when used for MI injection.

Table 2: Summary of proton drivers

| Parameter | MAP | JPARC RCS | JPARC MR |
|---|---|---|---|
| Injection E | 8 | 0.4 | 3 GeV |
| Top Energy | 8 | 3 | 30-50 GeV |
| Power | 4MW | 1MW | 0.67-1.1 MW |
| Frequency | 15 | 25 | 0.3 Hz |
| Emittance, 95%, N | 30π | 153π | 153π mm-mrad |
| Admittance | 50π | 200π | 300π |
| p/cycle | $2.1 \times 10^{14}$ | $8.4 \times 10^{13}$ | $3.5 \times 10^{14}$ |
| bunches | 4→1 | 2 | 8 |
| N/bunch | $5.2 \times 10^{13}$→$2.1 \times 10^{14}$ | $4.2 \times 10^{13}$ | $4.2 \times 10^{13}$ |
| kJ/bunch | 67→268 | 20 | 200→320 |
| Circumference | 308.2 | 348 | 1568m |
| Tune | 7.94/6.91 | 6.7/6.3 | 22.3/22.3 |
| $\gamma_t$ | 9.07 | 9.14 | i31.7 |
| Beam pipe R | 5 | 12.5 | 6.5cm |

## POTENTIAL UPGRADE PATHS

The JPARC RCS is close to space charge limits at design parameters and very large increases in intensity appear to be limited. An incremental improvement program up to ~1.5 MW is considered, utilizing improved orbit and betatron control to increase the transverse aperture, lower relative peak current (increased longitudinal painting), and increase of space charge tune shift. Further increases in allowed space charge could increase this to ~2MW. Higher power could possibly be obtained by increasing the injection energy. An increase to 800 MeV injection could increase the intensity to the ~4MW scale.

The upgrade path for the JPARC MR contains several possibilities. As discussed above, the present program can reach ~1.3 MW, by increasing intensity per bunch to $4.2 \times 10^{13}$ and rep rate to 1.25Hz at 30 GeV. The beam output energy could be increased to 50 GeV, but that would reduce the rep rate, and could increase total beam energy/bunch by 5/3, but not overall beam power. Intensity could be increased by increasing the injected intensity and/or injection energy. JPARC is considering an ~8 GeV linac injector as a possible future upgrade, and that would enable a multi MW intensity, possibly ~4 MW. The pulse structure would be compatible with UMS muon collider or neutrino factory applications.

An 8 GeV Booster Ring is also under consideration. This could be fed by a new linac or by 3 GeV injection from the RCS and could be a multi MW 8 GeV source itself, or a higher–energy injector for the MR, enabling higher power at 30—50 GeV.[9, 10]

## CONCLUSION

The current performance of the JPARC RCS and MR accelerators, with modest upgrades and operational variations, demonstrate that proton beam sources at the parameters needed for muon colliders and neutrino factories can be constructed. While perhaps not completely optimal (The output beam energy of the RCS is smaller and the energy of the MR is larger than the ~8—15 GeV optimum.), the JPARC facility itself could become the required proton source. JPARC experience also provides some guidance for the present Fermilab proton beam upgrade plan.